\documentclass{ws-nl}

\usepackage{multicol}
\usepackage{graphicx}
\usepackage[superscript]{cite}
\usepackage{color}

\newcommand{\bra}[1]{\left|#1\right>}

\newcommand{\mt}[1]{\text{#1}}
\newcommand{\fnal}[1]{[n(\vec{#1}\,)]}
\newcommand{\fn}[1]{(#1)}
\newcommand{\vl}{von Lilienfeld }
\newcommand{\tk}{Tkatchenko }

\begin{document}

\catchline{2}{2}{2012}{}{}
\markboth{B. Kolb and T.Thonhauser}{Molecular Biology at the Quantum Level}

\title{MOLECULAR BIOLOGY AT THE QUANTUM LEVEL:\\
CAN MODERN DENSITY FUNCTIONAL THEORY\\
FORGE THE PATH?}

\author{Brian Kolb and T. Thonhauser}
\address{Department of Physics, Wake Forest University,\\
1834 Wake Forest Road, Winston-Salem, NC 27109, USA\\
\email{kolbba5@wfu.edu, thonhauser@wfu.edu}}

\maketitle


\begin{abstract}
Recent years have seen vast improvements in the ability of rigorous
quantum-mechanical methods to treat systems of interest to molecular
biology.  In this review article, we survey common computational methods
used to study such large, weakly bound systems, starting from classical
simulations and reaching to quantum chemistry and density functional
theory. We sketch their underlying frameworks and investigate their
strengths and weaknesses when applied to potentially large biomolecules.
In particular, density functional theory---a framework that can treat
thousands of atoms on firm theoretical ground---can now accurately
describe systems dominated by weak van der Waals interactions.  This
newfound ability has rekindled interest in using this tried-and-true
approach to investigate biological systems of real importance. In this
review, we focus on some new methods within density functional theory
that allow for accurate inclusion of the weak interactions that dominate
binding in biological macromolecules. Recent work utilizing these
methods to study biologically-relevant systems will be highlighted, and
a vision for the future of density functional theory within molecular
biology will be discussed.
\end{abstract}


\begin{multicols}{2}

\section{Introduction}

The scientific disciplines (e.g.\ biology, chemistry, physics) once
stood well separated from each other, with practitioners from each
approaching different questions in different ways. These divisions are
beginning to blur, however, as answers to questions from one field
increasingly require techniques and knowledge built up in another.
There is evidence of this effect in the increasing need for
interdisciplinary collaborations to solve problems arising in distinct
fields. A particularly poignant example of this blurring of lines is the
field of molecular biology, where researchers try to build an
understanding of biological systems starting at the molecular level.
Concepts from chemistry and physics arise naturally in such endeavors
and this has bred a symbiotic relationship between biologists, chemists,
and physicists, who now seek to answer similar
questions.\cite{Deniz2008,Keller1990,Westerhoff2004}

Some of the most important questions arising in this arena relate to the
structure and function of biological
macromolecules.\cite{Wright1999,Muller2002} For example, for rational
drug design to be viable, a detailed knowledge of the interactions
between a target protein and a potential drug molecule is necessary to
understand whether the drug will bind to the protein at the right
location and in the right way.\cite{Verlinde1994} From there, an
atomic-level understanding of the protein itself is necessary to
understand how allosteric effects turn a drug binding event into a
change in the behavior of the protein.\cite{Cui2008} These details
cannot come from a \emph{top down} investigation of the molecules, nor
can they come from simply observing the changing behavior as a function
of drug binding. Part of the physics lies in the statistical mechanics
of protein conformations, and part resides in the communication networks
within the protein, the elucidation of which hinges on the detailed
physics of the binding event and the transmission of information from
the binding site to a possibly distant effector site.

The same can be said of the {}``holy grail'' of molecular
biology---understanding protein folding and how the structure of a
protein relates to its function.  Coarse-grained models provide some
insight into the process of protein folding\cite{Clementi2008}, but a
true understanding of the process and the ability to reliably predict
how a protein will fold requires an atomic-level understanding of the
interactions within a particular protein. There are many other examples
of the need for atomistic detail in molecular biology. Ultimately, all
properties of biological macromolecules---such as DNA, RNA,
proteins---are governed by minute details involving the atomic and
electronic structure of their constituent parts as well as the
interactions between neighboring pieces of the molecule. Even dynamic
conformational changes that may be essential to a particular process are
ultimately governed by these interactions and similar interactions with
the surrounding environment.

Developing an atomic-level understanding of large molecular systems is
not an easy task and, until recently, the application of accurate
quantum mechanical methods to such systems was infeasible.  This review
highlights recent advances made in the fields of computational physics
and physical chemistry that can aid in building such an understanding.
After discussing classical simulations and common quantum-chemistry
approaches, we focus specifically on advances within density functional
theory (DFT)---a framework used successfully for decades in the field of
condensed matter physics---which affords unprecedented accuracy and
utility in treating large, weakly bound molecular systems. These new
methods will be discussed and paired with a survey of their use on
biologically-relevant molecular systems. Possible future applications of
these methods will also be addressed.

\section{Survey of Common Computational Methods}
\label{sec_methods}

\subsection{Classical simulations}

For many purposes, the best present-day methods to study
biologically-relevant systems are classical force field models. Such
methods allow one to study the large-scale dynamics of systems with
perhaps millions of atoms over biologically-relevant
timescales.\cite{MacKerell2004} This is by far the most common
computational method of study for macromolecules, and has provided
indispensable insight into numerous biological systems.

The main goal of a force field is simple; to represent the energy and
forces of a collection of atoms using a physically-motivated, yet
relatively straight-forward, algebraic expression. This simplicity is
what allows the simulation of large systems over significant
timescales.\cite{MacKerell2004} Generally, the physical motivation for
terms in the energy Hamiltonian come from macroscopic physics. For
example, many force fields treat bond stretches and angle flexes as
classical harmonic oscillators obeying Hooke's law. This is, in fact,
what is meant by the phrase \textit{classical} force field.

In its most basic form, a typical force field can be written as a sum of
separate contributions to the total
energy,\cite{MacKerell2004,Oostenbrink2004,Rappe1992,Halgren1995,MacKerell2001}
i.e.\
\begin{eqnarray}
E_{\rm ff} &=& E_{\rm bonds}+E_{\rm angles} +
   E_{\rm dihedrals}+E_\text{non-bonded}\nonumber\\[1ex]
   &=& \sum_{b}{\frac{1}{2}k_{b}(d_{b}-d_{0})^{2}}\nonumber\\
   &+& \sum_{a}{\frac{1}{2}k_{a}(\theta_{a}-\theta_{0})^{2}}\nonumber\\
   &+& \sum_{d}{\frac{1}{2}k_{d}\big[1+\cos(n\phi-\delta)\big]}\nonumber\\
   &+& \sum_{nb}{\bigg[\frac{q_{i}q_{j}}{r_{ij}}+
   \bigg(\frac{C_{12}}{r_{ij}^{12}}-
   \frac{C_{6}}{r_{ij}^{6}}\bigg)\bigg]}\;.\label{E_FF}
\end{eqnarray}
The first term on the right hand side of Eq.~(\ref{E_FF}) represents an
harmonic oscillator (with spring constant $k_b$) in bond length between
each pair of covalently-bonded atoms within the system. The second does
the same for the three-atom angle term.  Dihedral angles are treated
with a fairly shallow periodic potential, represented by the third term.
The last line of Eq.~(\ref{E_FF}) represents non-bonded interactions and
includes a coulomb term for charge-charge interactions and a
Lennard-Jones (6--12) potential to account for van der Waals type
interactions.  Some force fields add additional terms for out-of-plane
motions (improper dihedrals) or higher-order terms.\cite{MacKerell2004}
Variations on the functional form are also sometimes applied.  For
example, a Morse potential can be used in place of the harmonic bond
term to allow for bond breaking during a
simulation.\cite{Rappe1992,MacKerell2004} 

For all their usefulness, so called ``Class I'' force field approaches
suffer from some drawbacks. First, treating microscopic phenomena using
macroscopic theory is, in essence, a mean-field approach. The
quantum-mechanical interactions between electron clouds are averaged
over. This, along with the assumed form for all physical interactions,
does not allow new physics to be uncovered. The only physics present in
the simulation is what was explicitly included, meaning one cannot gain
any true atomic-level insight into the underpinnings of interesting
phenomena.  Second, the simplicity of the mean-field approach used in
force field simulations means that they are generally incapable of
transferably achieving chemical accuracy.\cite{Hagler1994} While bulk
motions and general trends can often be gleaned from such simulations,
the precise movements and behavior of atoms are probably not accurate.
This poses a significant problem for applications such as drug design,
where one seeks to find a small molecule (an enzyme inhibitor perhaps)
that binds with a certain affinity to a site in the protein.

Biophysicists and biochemists have already made substantial headway
against this problem.\cite{Hagler1994} Originally, force fields included
the partial charge on an atom as a fitting parameter.  The charge was
assumed fixed during the simulation so effects of polarization could not
be treated. The next generation of force fields incorporates the ability
of charge to rearrange during a simulation. Such \emph{polarizable}
force fields incorporate some of the quantum effects necessary to
accurately model molecular systems. One example of this new type of
force field is the AMOEBA force field, which includes both static and
dynamic polarizabilities and represents a significant step towards
accurate energetics from a force field.\cite{Ponder2010} In addition,
newer force fields often include cross-terms that account for how
changes in one internal coordinate affect other energy terms. These help
improve accuracy and transferability but cannot correct for the lack of
an explicit quantum mechanical treatment.

\subsection{Incorporating quantum mechanics}

The obvious solution to the shortcomings of the classical force field
methods is to directly include quantum mechanics in calculations.
Therefore, the solution to the problem is straight-forward; one simply
has to solve the time-independent Schr\"odinger equation
\begin{equation}
\hat{H}\left|\Psi\right>=\varepsilon\left|\Psi\right>\,,
\end{equation}
where the Hamiltonian  $\hat{H}$ in atomic units is given by
\begin{eqnarray}
\hat{H} &=& -\frac{1}{2}\sum_{i}^{n}\nabla_{i}^{2}
-\sum_{i,J}\frac{Z_{J}}{|\vec{r}_{i}-\vec{R}_{J}|}\nonumber\\
&& {}+\frac{1}{2}\sum_{i\ne j}\frac{1}{|\vec{r}_{i}-\vec{r}_{j}|}
+\frac{1}{2}\sum_{I\ne J}\frac{Z_{I}Z_{J}}
{|\vec{R}_{I}-\vec{R}_{J}|}\,.\label{Schrodinger}
\end{eqnarray}
Here, lower-case letters represent electronic degrees-of-freedom,
upper-case letters represent nuclear degrees-of-freedom (including
charge Z), $\varepsilon$ is the energy of the system, and the explicit
representation of the Hamiltonian $\hat{H}$ follows from specialization
to an isolated system of atoms under the Born-Oppenheimer approximation.

The unknown function $\bra{\Psi}$ is the wave function for the electrons
and from it (along with knowledge of the nuclear positions
$\{\vec{R}_{I}\}$) one can calculate all accessible properties of the
system. Unfortunately, $\bra{\Psi}$ depends on the coordinates of all
electrons within the system and, as a result, direct solution of
Eq.~(\ref{Schrodinger}) for the full many-body wave function is
difficult or impossible for all but the most trivial systems. For
example, a single neutral water molecule has 10 electrons, so its wave
function is a function of $30$ variables (i.e.\ 10 electron positions in
three dimensions). While an analytical solution in this simple case is
already not possible, it is conceivable that the Schr\"odinger equation
could be solved numerically. However, to store the wave function on a
numerical grid consisting of 10 points in each dimension (a laughably
coarse grid) using single precision numerics would take $4\times10^{30}$
bytes (approximately $10^{18}$ TB) of storage. This is {}``the curse of
dimensionality'' on a grand scale and renders full solution of the
Schr\"odinger equation for most systems utterly intractable.

Surmounting this fundamental problem in a physical way is not easy and
has consumed the efforts of chemists and physicists alike for decades.
From those efforts, however, have sprung a number of useful approaches.
These can be split into two categories, wave function theories and
density functional theory, both of which we will discuss in detail
below.

All the methods described in what follows have exhibited great success
in describing various quantum-mechanical properties of molecules and
materials in general.  However, when dealing with biologically-relevant
systems two special considerations arise: (i) such systems are typically
quite \emph{large} and (ii) their structure and binding is often
dominated by weak \emph{van der Waals} interactions.  Since this review
is focused on biological applications of quantum-mechanical methods,
special attention will be paid to the ability of each method to scale
well with system size and to adequately describe van der Waals
interactions. As such, the ability of a method to treat large systems
involving van der Waals interactions will determine its applicability to
the biologically-relevant systems considered here.

\subsection{Wave function approaches}

A simple solution to the dimensionality problem introduced by
Eq.~(\ref{Schrodinger}) is to seek solutions of the form
\begin{equation}
\label{psitophi}
\Psi(\{\vec{r}_i\})=\phi_1(\vec{r}_1)\,\phi_2(\vec{r}_2)
\cdots\phi_n(\vec{r}_n)\;,
\end{equation}
that is, to assume that the total electron wave function can be
separated and written as a product of single-electron states (orbitals)
$\phi_i$. The Pauli-exclusion principle and the anti-symmetry of the
wave function can be enforced by forming a Slater determinant of the
single-particle solutions. Since the Fock operator used to find the
orbitals depends explicitly on those orbitals, the resulting equations
are generally solved self-consistently. This approach is a form of
mean-field theory where each electron responds to the average field
created by all other electrons residing in their single-particle
orbitals.  The advantage of this approach is that each orbital is now a
function of the three spatial coordinates, making numerical calculations
computationally feasible. Wave function methods are described in detail
in Ref.~\refcite{Szabo1996}.

The Hartree-Fock (HF) method, which takes this approach, is relatively
fast and based on sound quantum mechanics, but the approximations
invoked by its use miss some crucial physics. In particular, electrons
are dynamic entities. The total energy of the system can be lowered if,
averaged over some degree-of-freedom, the electrons correlate their
behavior. Correlation is (almost) completely missed in the Hartree-Fock
method, which explicitly assumes single-particle states---the static
correlation due to the Pauli exclusion principle is fully accounted for.
Nevertheless, HF theory is a good first-order starting point for
corrections that incorporate electron correlation into the total wave
function and its associated energy. Such methods are termed post-HF
methods, since they use the results of a HF calculation as a starting
point to incorporate electron correlation explicitly.

There are many post-HF methods that exhibit various accuracies coming at
related computational costs. One of the best features of the wave
function methods is their segregation into a hierarchy of so called
{}``levels of theory''. Thus, one knows in some sense, to what degree a
result can be trusted, depending on the precise method used. If better
results are desired, one merely has to progress to a higher level of
theory. Basis sets (the set of functions used to expand the wave
functions) are also of critical importance. They too, however, exhibit a
hierarchy of complexity and applicability.  Figure~\ref{fig_hf} gives a
cartoon depiction of how one can approach the numerically exact solution
$|\Psi\rangle$ by combining a large basis set with a high level of
theory.

\begin{figurehere}
\vspace{4ex}
\centerline{\includegraphics[width=0.8\columnwidth]{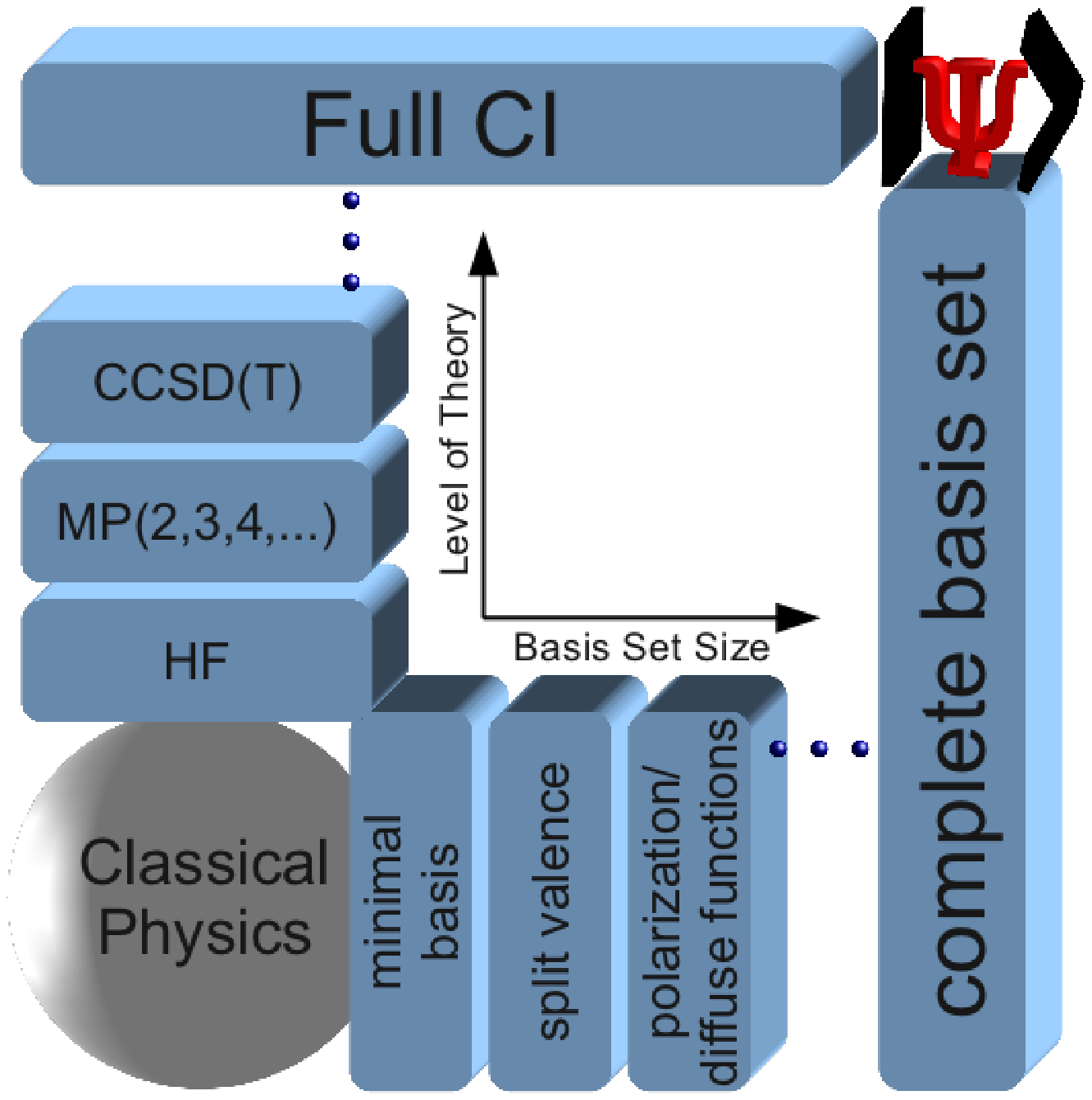}}
\caption{\label{fig_hf}Map of the route from classical physics to
quantum physics via quantum chemistry. Basis sets are represented on the
horizontal axis and increase in size as more functions are added. The
level of theory is indicated on the vertical axis. There is a
concomitant increase in computational complexity as one moves along the
path from classical physics to quantum physics. A plot of this nature
is often called a \emph{Pople diagram}.}
\end{figurehere}

The most rigorous method to include electron correlation is full
configuration interaction (CI). In the CI method, one starts as usual
with the orbitals found by a Hartree-Fock calculation. Instead of using
a single Slater determinant of these functions, however, a linear
combination of Slater determinants is formed, each one corresponding to
one possible ordering of electrons in the orbitals. In other words, all
possible combinations of electron excitations are given a Slater
determinant, and the optimized linear combination of these yields the
numerically exact wave function.  This renders Full CI a combinatorial
problem---taking a given number of electrons and producing all possible
excitations to a given set of orbitals. Thus, full CI scales factorially
with the number of basis functions used and therefore is not practical
in all but the smallest of systems.

Perhaps the next best thing to a full CI calculation is to use coupled
cluster theory. The coupled cluster approach mimics CI but using only
small numbers of electron excitations, usually considering only
excitations of one to three electrons. The most common variant of
coupled-cluster theory is notated CCSD(T), which includes single and
double excitations iteratively, and triple excitations perturbatively.
This has proven incredibly reliable and represents the {}``gold
standard'' for accurate quantum-chemistry calculations. Although it has
polynomial (rather than factorial) scaling in the number of basis
functions used, the asymptotic scaling of $\mathcal{O}(N^{7})$ for the
generally used form renders this approach mainly useful on relatively
small systems of perhaps 30--50 atoms.

Among the most used post-HF methods is M{\o}ller-Plesset perturbation
theory at second order (MP2) or higher order (MP3, MP4, \dots).  In
perturbation theory, one seeks to find the solution of
\begin{equation}
\hat{H}\bra{\psi}=\Big[\hat{H_{0}}+\lambda\hat{H'}\Big]
\bra{\psi}=E\bra{\psi}\,,\label{perturbation}
\end{equation}
where the perturbation strength factor $\lambda$ is assumed small, and
the solution to the unpertured problem ($\lambda=0$) is already known.
In this case, $\hat{H}$ is the non-interacting Hartree-Fock Hamiltonian
and $\hat{H'}$, which is assumed to be small in effect relative to
$\hat{H_{0}}$, is the Hamiltonian for inclusion of electron correlation.
MP2 expands this expression in terms of powers of $\lambda$ up to second
order. This can be used to correct both energies and wave functions.

MP2 has shown great success, but it is not perfect. Comparison with
coupled cluster and full CI methods have shown that MP2 often
significantly overestimates the correlation, especially in delocalized
$\pi$ systems.\cite{Schwabe2009,Grimme2004} Usage of the higher-order
expansions (e.g. MP4) may yield increased accuracy, but the results are
not as straightforward as one might hope, as convergence of the
M{\o}ller-Plesset series has been shown to be unreliable.\cite{He2000}
In many cases, estimates of correlation may get worse with increasing
order; sometimes oscillating or even diverging in the worst
cases.\cite{Leininger2000} Convergence depends on both the system under
study and the basis set being employed, with poor results often
accompanying use of the diffuse functions required to correctly model
dispersion interactions. Nevertheless, MP methods are highly prized in
quantum-chemistry wave function calculations because they contain a good
balance of accuracy and computational efficiency. The asymptotic scaling
of MP2 (as $\mathcal{O}(N^5)$) makes it substantially cheaper than
high-level coupled cluster methods. MP2 can be used on systems of
respectable size. A system with a hundred atoms or more is not out of
the reach of an MP2 calculation on a high-end computer.

\begin{figurehere}
\vspace{4ex}
\centerline{\includegraphics[width=0.8\columnwidth]{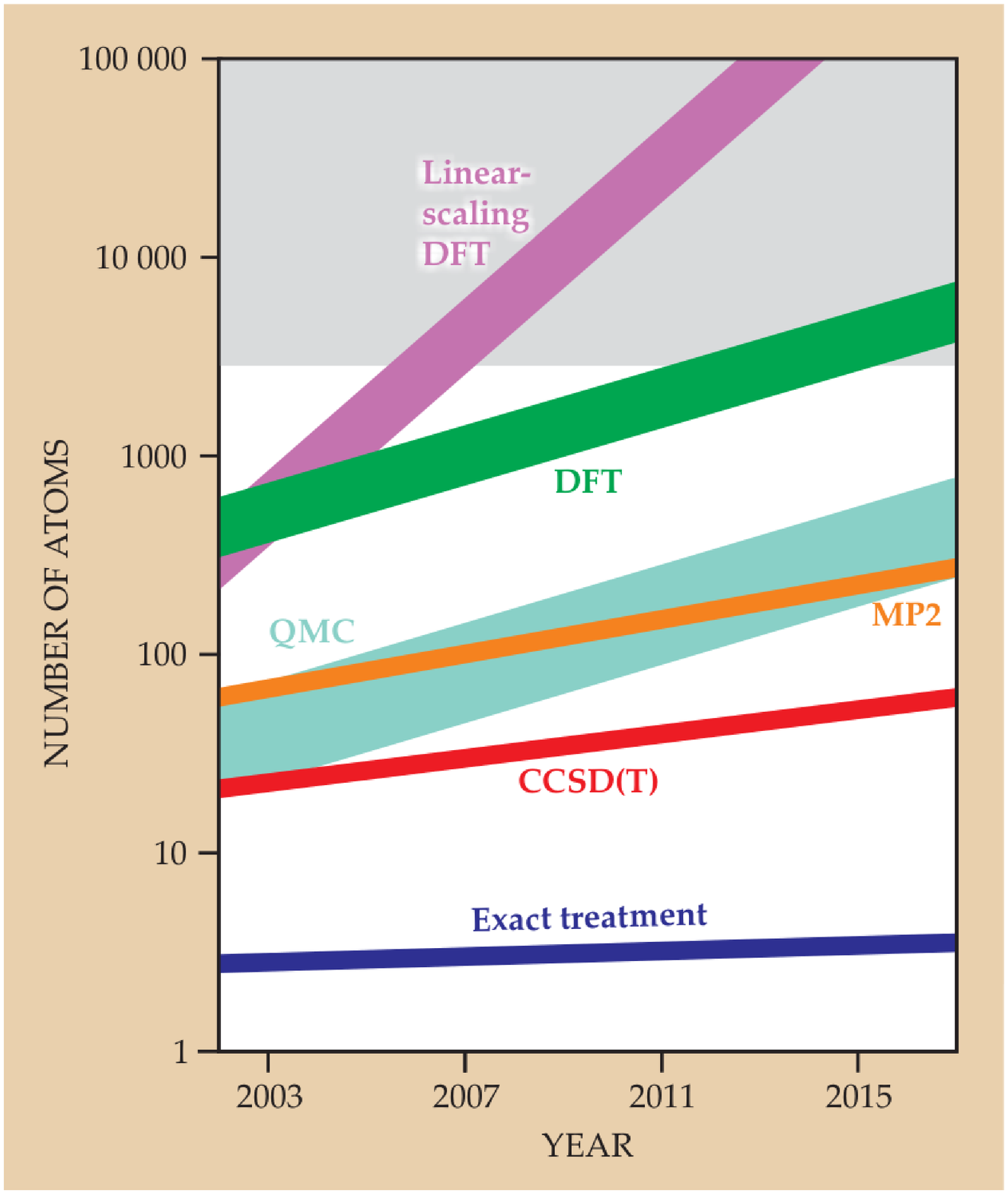}}
\caption{\label{scaling}Maximal system size (measured by number of
atoms) that various quantum mechanical methods can treat, as a function
of time. ``Exact treatment'' refers to an exact solution to the
Schr\"odinger equation and QMC stands for Quantum Monte Carlo (not
discussed here). All other methods are discussed throughout the text.
(Reprinted with permission from
Ref.~[\protect\refcite{Head-Gordon2008}]; \copyright\ 2008 American
Physical Society).}
\end{figurehere}

\subsection{Density functional theory}
\label{sec:DFT}

Wave Function theories have a number of nice properties, but they scale
poorly with systems size. A completely different approach, density
functional theory (DFT), scales as $\mathcal{O}(N^3)$, and is therefore
much more amenable to calculation of large systems. Calculations can be
performed on systems consisting of perhaps several thousand atoms,
making it applicable to biochemical systems.

In 1964 Hohenberg and Kohn\cite{hohenberg1964} published a seminal paper
showing that the quantum-mechanical energy of a set of atoms can be
written uniquely as a functional of the electron charge density (within
the Born-Oppenheimer approximation). Furthermore, the charge density
$n_0(\vec{r\,})$ that minimizes this functional is the ground-state
charge density for the system, and all measurable properties of the
system can be written in terms of this optimal charge density. This
avoids the dimensionality problem of Eq.~(\ref{Schrodinger}) by shifting
the quantity of interest to the charge density in real space, a function
of only three variables regardless of the number of electrons.

Density functional theory as a modern approach was initiated when Kohn
and Sham\cite{kohn1965} wrote the energy as a density functional of the
form
\begin{eqnarray}
\label{DFT}
E_{\rm DFT}\fnal{r} &=& E_{\mt{k}}\fnal{r}+E_{\mt{N-e}}\fnal{r}+E_{\mt{e-e}}\fnal{r}\nonumber\\
&+& E_{\mt{xc}}\fnal{r}+E_{\mt{N-N}}\;,
\end{eqnarray}
where $E_{\mt{k}}$ is the total kinetic energy of the system, in
principle written as a density functional, but in practice written as a
functional of the Kohn-Sham orbitals. An analytical density functional
for $E_{\mt{k}}$ is not known, but approximations to it lead to so
called orbital-free methods. The final term in Eq.~(\ref{DFT}) is the
nucleus-nucleus repulsion term, which can be treated as a simple
additive constant since it is uniquely determined by the positions of
the nuclei and these are decoupled from the quantum-mechanical problem
by use of the Born-Oppenheimer approximation. Analytical expressions are
known for both $E_{\mt{N-e}}$ (the nucleus-electron, effective 1-body
term) and $E_{\mt{e-e}}$ (the Hartree term giving the average
electron-electron interaction), leaving the
\textit{exchange-correlation} functional $E_{\mt{xc}}\fnal{r}$ as the
sole unknown object in Eq.~(\ref{DFT}).

If the exchange-correlation functional and its \textit{functional}
derivative with respect to the density were known, they could be used to
optimize the total energy functional ($E_{\rm DFT}\fnal{r}$) with
respect to the density, thereby finding the ground-state density.
Since, $E_{\mt{xc}}$ is not known however, it must be approximated in
some way. This is the main approximation in DFT and determines the
method's applicability to a particular system. Not surprisingly then,
much effort is put into improving the approximations made in generating
$E_{\mt{xc}}\fnal{r}$.

One approach is to assume that the exchange-correlation energy is a
\textit{local} functional of the density---one that depends on
$n(\vec{r}\,)$ in a point-wise fashion.\cite{perdew1981,ceperley1980}
This local density approximation (LDA) is good when the density is
slowly varying, becoming exact in the limit of a uniform electron
density. 

Despite its simplicity, the LDA is amazingly good in many systems,
especially in those with relatively concentrated charge density such as
crystalline environments where metallic or covalent bonding dominates.
For molecules, where directional covalent bonds are the primary
interaction, it tends to perform less adequately. One can imagine the
LDA as the zeroth-order term in the Taylor expansion of the density
about each point, and envision adding additional, derivative-dependent
terms.  A functional depending on the density and its gradient (first
derivative) in a point-wise fashion is called a semi-local functional,
and the approximation of the true energy functional in this way is
called the generalized gradient approximation (GGA).\cite{perdew1996}
This approximation is a substantial improvement over LDA in many
systems, particularly molecules.

The $E_{\mt{xc}}$ term in Eq.~(\ref{DFT}) must approximate the effects
of both exchange, which removes the unphysical electron self interaction
while enforcing the Pauli exclusion principle, and correlation, which
roughly speaking, accounts for the fact that each electron experiences a
highly dynamic environment rather than a \textit{mean field} of the
other electrons. In Hartree-Fock theory, the form of the exchange
operator is known, so exchange could be treated exactly and combined
with an approximate correlation functional. Unfortunately, most
functionals exhibit serendipitous error cancellations between their
exchange and correlation pieces, making just using the correlation
contribution prone to large errors. In 1993 Becke proposed using a
50\%--50\% mix of exact exchange and LDA,\cite{Becke1993} eventually
leading to the 3-parameter B3LYP functional\cite{Stephens1994} and
similar hybrid functionals, which are among the most accurate
functionals for covalently-bound molecules. Early successes of hybrid
functionals led some to erroneously believe that they could describe
weak van der Waals interactions.  Unfortunately, hybrid functionals,
being a linear combination of exact exchange and (semi)-local
exchange-correlation approximations, cannot account for van der Waals
interactions.  This is because van der Waals interactions are a
non-local correlation effect, and any functional that is local or
semi-local in correlation is---by construction---not able to reliably
describe them.\cite{RevModPhys.82.1887} There is ample discussion in the
literature of the poor performance of standard hybrid functionals in
weakly bound
complexes.\cite{Zhao2005,Ortmann2008,Rutledge2010,Huenerbein2010,Thanthiriwatte2011}

\subsection{van der Waals interactions in DFT}

As evident from the discussion in Section~\ref{sec:DFT} and
Fig.~\ref{scaling}, DFT is capable of treating large systems of perhaps
thousands of atoms, which is one of the requirements if it is to be
applicable to systems of interest in molecular biology. However, at the
same time, it also has to be able to accurately describe weak van der
Waals interactions, which play an important role in biomolecules.
Historically, DFT has not performed well when applied to systems with
van der Waals interactions---this is probably the single most important
problem that has prevented DFT from gaining a strong foothold in
molecular biology. Below we discuss the shortcomings of standard DFT and
several recent developments that overcome this barrier, leading to a
full applicability of DFT to large biomolecules.

In standard DFT, the exchange-correlation functional is often assumed
to be \emph{local}, i.e.\ a single spacial integral of the
exchange-correlation energy density, which depends explicitly on the
charge-density.  This approach leads to the so called local density
approximation (LDA). Adding a dependence on the gradient of the charge
density results in the generalized gradient approximation (GGA), while
inclusion of higher-order derivative terms yield meta-GGA
functionals. However, this approach fails to correctly account for van
der Waals (vdW) interactions, which are non-local correlation effects;
they occur between physically separated regions of charge, generally
with little overlap of their density functions. Capturing these
effects correctly requires a functional that expresses the
exchange-correlation energy as (at minimum) a double spacial integral.
van der Waals interactions, ubiquitous in polyatomic systems, occur
when electron motions in one atom (or within one molecule) correlate
with electron motions in a nearby atom (or molecule) setting up
transient but interacting multipoles within each.\cite{Grimme2011}
Correlation between electrons lowers their energy relative to
uncorrelated electrons, so the van der Waals force is always
attractive. In some systems, crystalline NaCl for example, the
contribution of these interactions to the overall binding are
negligible. In other systems these interactions can be an appreciable
part of the overall interaction. Nobel gas dimers such as Ar$_{2}$ and
Kr$_{2}$ are held together entirely by van der Waals interactions.
Large diffuse molecular systems (prime examples being bio\-logical
macromolecules) rely quite heavily on van der Waals interactions for
their stability, so such interactions play an integral role in their
behavior.

With this in mind, numerous attempts were made to include the ability to
capture van der Waals interactions within conventional DFT. A thorough
account of all these efforts is beyond the scope of the present review,
but several promising approaches will be discussed.

\subsection{DFT-D}

As stated earlier, van der Waals interactions arise when electronic
motions within separated atoms correlate, setting up transient multipole
moments within the individual atoms. One can expand the dispersion
energy of two arbitrary, polarizable charge densities in terms of the
interactions of induced multipoles. If a point of interest is located at
a distance $r$ that is large compared to some characteristic length
scale of the charge distribution, the \emph{pairwise} dispersion energy can be expanded
in powers of $1/r$ as\cite{VonLilienfeld2010}
\begin{equation}
\label{expansion}
E_{\mt{disp}}=-\frac{C_{6}}{r^{6}}-\frac{C_{8}}{r^{8}}
-\frac{C_{10}}{r^{10}}-\cdots\,,
\end{equation}
where the constants $C_{i}$ correspond to a particular system and
determine the relative strengths of the various terms.

For sufficiently large distances $r$, the dipole-dipole term dominates
and dispersion interactions go as 1/$r^{6}$. This observation is the
basis of the density functional theory with added dispersion (DFT-D)
method.\cite{Grimme2004,Grimme2006,Grimme2007} Typically, this method
works by adding to the total energy a pairwise atomic correction of the
form
\begin{equation}
\label{DFTD}
E_{\rm vdW}=-\frac{1}{2}\sum_{I\ne J}f_{\mt{damp}}\left(R_{IJ}\right)\frac{C_{6}^{IJ}}{R_{IJ}^{6}}\;,
\end{equation}
where $E_{\rm vdW}$ is the dispersion energy, $C_{IJ}$ is an
empirically-derived coefficient that is atom-pair-dependent,
$f_{\mt{damp}}\left(R_{IJ}\right)$ is a damping function, and the sum
runs over all pairs of atoms. The damping function ranges from 0 at
small $R_{IJ}$ to 1 for larger separations, and is required because the
asymptotic 1/$R^{6}$ form becomes unphysical as distances become small.
The specific form of the damping function plays a role in the accuracy
of the technique.\cite{halgren1992,liu2009a} Too weak a damping with
decreasing distance results in over-counting of the interaction energy.
Too strong a damping will weaken the vdW interactions at relevant
ranges. The most critical aspect of the damping function is how it
behaves at intermediate distances near the bonding length of a vdW bond.
Much attention has been paid to the form of the damping function, and
opinions differ on its optimal form. Commonly, the damping function is
given the
form\cite{Wu2002,Zimmerli2004,Grimme2011c,liu2009a,Tkatchenko2009}
\begin{equation}
f_{\mt{damp}}=\frac{1}{1+e^{-\alpha\big(\frac{R_{IJ}}{R_{0}}-1\big)}}\;,
\end{equation}
where $\alpha$ is a chosen constant and $R_{0}$ sets the relevant
distance scale for the interaction of atoms $I$ and $J$ and is generally
chosen to be the sum of their van der Waals radii.  This was the form
chosen by Grimme in 2004\cite{Grimme2004}, when he published a set of
$C_{6}$ coefficients based on a database of dipole oscillator strength
distributions, for a number of important atoms and demonstrated the
method's effectiveness on a large set of molecular systems. The values
of the $C_{6}$ coefficients (and corresponding vdW-radii) depend
somewhat on the choice of exchange-correlation functional, so Grimme
added an empirical parameter he called $s_{6}$ that scales the
interaction, adjusting its strength to the functional being used.
Approaches like the DFT-D method are not new, dating back at least as
far as London himself\cite{london1930}, but they have proved extremely
useful at many levels of atomic theory, and continue to be so within
DFT.

\subsection{DFT+vdW}
\label{DFT+vdW}

The pairwise dispersion correction given by Eq.~(\ref{DFTD}) is not a
density-functional, but instead relies on fitting to a chosen set of
external data. The data used in the fit and the interaction between the
dispersion correction and the functional coupled with it both affect the
results obtained. Such a fitting procedure can limit transferability
between systems. The original DFT-D approach of Grimme has been
re-parameterized many times both for improved accuracy and for
application to other systems.\cite{Grimme2004,Grimme2006,Grimme2010}

To improve on the transferability and overall accuracy of DFT-D,
Tkatchenko and Scheffler proposed an alteration, which uses a relative
$C_{6}$ coefficient calculated on-the-fly from the charge
density.\cite{Tkatchenko2009} In this approach (hereafter referred to as
DFT+vdW) they define the effective volume for an atom $A$ within a
system, relative to the free-atom volume as:
\begin{equation}
\label{Veff}
\frac{V_{A}^{\text{eff}}}{V_{A}^{\text{free}}}=
\frac{\displaystyle \int{r^{3}f_{A}\,(\vec{r}\,)\,
n(\vec{r}\,)\,d^{3}\vec{r}}}{\displaystyle \int{r^{3}\,
n_{A}^{\text{free}}(\vec{r}\,)\,d^{3}\vec{r}}}\;,
\end{equation}
where $f_{A}(\vec{r}\,)$ is the fraction of the density at $\vec{r}$
arising from atom $A$ in a linear combination of free-atom charge
densities.

Starting from the free-atom coefficients $C_{6}^{\text{free}}$, they
define the effective coefficient for an atom within a molecule or solid
as
\begin{equation}
C_{6}^{\text{eff}}=\bigg(\frac{V^{\text{eff}}}{V^{\text{free}}}\bigg)^{2}
C_{6}^{\text{free}}\;,
\end{equation}
with the hetero-nuclear combination rule for atoms $A$ and $B$ defined
as:
\begin{equation}
\label{comborule}
C_{6}^{AB}=\frac{2C_{6}^{AA}C_{6}^{BB}}{\frac{\alpha_{B}}{\alpha_{A}}C_{6}^{AA}+
\frac{\alpha_{A}}{\alpha_{B}}C_{6}^{BB}}\;,
\end{equation}
where $\alpha_{i}$ is the static polarizability of atom $i$. Thus, in
the DFT+vdW approach one may write
\begin{align}
\label{eq:c6}
E_{\rm vdW}[n(\vec{r}\,)]=-\frac{1}{2}\sum_{I\ne J}
\frac{C_{6}^{IJ}[n(\vec{r}\,)]}{R_{IJ}^{6}}\;,
\end{align}
with $I$ and $J$ ranging over all atoms. That is, the dispersion energy
can be written as a functional (albeit a non-universal one that depends
on the arrangement of nuclei) of the charge density.  Writing the
$C_{6}$ coefficients as density functionals allows for the
polarizability of atoms to be a dynamic, environment-dependent quantity.
If it could be calculated, the functional derivative of this expression
with respect to the charge density would yield the Kohn-Sham potential
for the dispersion energy, allowing the latter to be calculated
self-consistently. It is not clear at present whether this would
significantly affect the interaction energies of vdW compounds when
using the DFT+vdW method. Tkatchenko and Scheffler do note, however,
that the use of Eq.~(\ref{Veff}) largely cancels the charge density
differences arising from the use of different functionals, making their
method less sensitive to the particular exchange-correlation functional
used compared with static $C_{6}/r^{6}$ approaches.\cite{Tkatchenko2009}

In 2008, Tkatchenko and von Lilienfeld noted that many body effects can
play a significant role in the energetics of vdW-rich
systems.\cite{Tkatchenko2008} This is especially true in bulk, where
close-packed atoms can be geometrically arranged in many complex ways.
In particular, three-body, triple-dipole interactions can contribute
substantially to binding energies, typically raising them relative to
pure pairwise interactions. Given the recent surge of inquiry into metal
organic frameworks and other molecular crystals,\cite{Yanpeng2012} the
ability to account for this fact may become of increasing interest. In
2010, an expanded version of DFT+vdW was described, wherein an
additional three-body term was added.\cite{VonLilienfeld2010} The
three-body term was given the form
\begin{equation}
E_{\rm vdW}^{\text{3-body}}=C_{9}^{ABC}\frac{3\cos(\phi_{AB})\cos(\phi_{BC})
\cos(\phi_{AC})+1}{R_{AB}^{3}R_{BC}^{3}R_{AC}^{3}}\;,
\end{equation}
where $R_{IJ}$ is the distance between atoms $I$ and $J$ and $\phi_{IJ}$
is the angle between $\vec{R}_{KI}$ and $\vec{R}_{KJ}$.  The
formulation of this expression into a density functional then follows in
a fashion similar to that of the $C_{6}$ term, the fundamental
difference being the inclusion of an angular dependence in the damping
function.

\subsection{vdW-DF}

An alternative approach to the addition of pairwise atomic dispersion
terms is to express the total energy of a system directly as a
\textit{non-local} functional of the density. That is, to write the
exchange-correlation functional in such a way that it depends
simultaneously on the charge density at multiple points. In principle,
this is the optimal approach because the true exchange-correlation
functional is \emph{fundamentally} a non-local functional.  Treating it
on such a footing allows for its integration into DFT in a seamless and
self-consistent manner.

In the van der Waals density functional (vdW-DF) approach, the
exchange-correlation functional $E_{\mt{xc}}$ takes the form
\begin{eqnarray}
\label{nl}
E_{\rm xc} &=& E_{\mt{xc}}^{\mt{local}}+E_{\mt{xc}}^{\mt{non-local}}\\
   &=& E_{\mt{xc}}^{\mt{local}}+\int n\fn{\vec{r}_1}\,\phi(\vec{r}_1,\vec{r}_2)
   \,n\fn{\vec{r}_2}\,d^3\vec{r}_1d^3\vec{r}_2\;,\nonumber
\end{eqnarray}
where $E_{\mt{xc}}^{\mt{local}}$ is a local-like piece of the functional
that is assumed to be well modeled by standard functionals and
$E_{\mt{xc}}^{\mt{non-local}}$ is a non-local piece, which is evaluated
by considering all \textit{pairwise} points in the charge density. The
kernel function $\phi$ describes how charge densities at $\vec{r}_{i}$
and $\vec{r}_{j}$ correlate.

A meaningful form for $\phi$ was described by Dion et al. in 2004,
leading to the van der Waals density functional (vdW-DF).\cite{dion2004}
This functional evolved from a less general one restricted to planar
geometries.\cite{Rydberg2003a} The analytical form of $\phi$ and its
numerical computation are onerous, but since $\phi$ itself does not
depend on the density, it can be calculated and tabulated
once-and-for-all.  The functional derivative of Eq.~(\ref{nl}) with
respect to the charge density was given in 2007,\cite{Thonhauser2007}
allowing for completely self-consistent calculation of energies and
forces using this method.

The functional as originally proposed required the evaluation of a
double integral over three-dimensional space, as one might expect from a
non-local functional. This made the use of the functional costly
relative to other local or semi-local options. However, in 2009
Rom\'{a}n-P\'{e}rez and Soler effected a great simplification, by
transforming the double integral into a single integral over Fourier
transforms using the convolution theorem.\cite{soler2009} Since Fourier
transforms are efficiently obtained and/or readily available in
plane-wave DFT codes, this dropped dramatically the time required to
evaluate the vdW-DF functional and made the cost of its use on par with
that of a similar GGA calculation.

It was quickly noted that, when used on vdW-rich systems, the functional
produced binding distances that were slightly larger compared with
experiment or high-level calculations.\cite{Thonhauser2006} This led to
the assertion that the revised Perdew-Burke-Ernzerhoff exchange
functional,\cite{Zhang1998} originally chosen to accompany the vdW-DF
because it exhibited minimal spurious binding of its own, was too
repulsive.\cite{Cooper2010} Lee et al.  revised the approach in 2010,
recommending the use of a less repulsive revised version of the
Perdew-Wang\cite{perdew1986,murray2009} exchange functional and changing
the value of a gradient coefficient.\cite{lee2010} These small changes
improved the method's accuracy for both energy and geometry in many
systems. For an in-depth review of this approach see
Ref.~\refcite{Langreth2009}.

\subsection{Other methods}

There are a number of other approaches that are capable of describing
van der Waals interactions within a DFT framework. A full listing is
beyond the scope of this review, but several of the more common
approaches are briefly discussed here.

In symmetry adapted perturbation theory (SAPT), the interaction energy
of a system is written as a perturbative expansion in terms of
physically-meaningful interactions. The Hamiltonian for a superposition
of non-interacting monomers is taken as $\hat{H}$$_{0}$, with the
interaction between monomers forming the perturbing
potential.\cite{jeziorski1994} Terms in the perturbation generally
include electrostatic, exchange, induction, and dispersion
interactions.\cite{Misquitta2005,Patkowski2006} The principle advantage
of this approach is that the relative contributions from different
physical interactions can be determined explicitly.  This leads to an
intuitive interpretation of the interaction energy.  The downside to the
approach is its computational cost since the method scales as
$\mathcal{O}(N^6)$ (when taken to second order) with increasing system
size.\cite{Misquitta2005} It is therefore limited to relatively small
systems. See Ref.~\refcite{jeziorski1994} for an excellent overview of
SAPT and its applications.

Zhao and Truhlar have developed a series of functionals designed to
obtain accurate energies for weakly-bound
systems.\cite{zhao2004,Zhao2006} These functionals have been shown to
work well for the $\pi$-stacking and hydrogen bonding interactions that
are omnipresent in biological
macromolecules.\cite{Zhao2005,Zhao2007,Hobza2008,Burns2011} The
advantage of these functionals is their efficiency, being essentially
the same computational cost as a typical DFT calculation. There are
several families of these functionals, designed and parameterized to
apply to different chemical situations. The functionals are known to
poorly describe dispersion interactions in the asymptotic limit, where
they decay exponentially rather than as $1/r^{6}$.\cite{Zhao2007}
Nevertheless, they have seen heavy use recently for their ability to
accurately and efficiently capture the short-ranged contributions to
dispersion interactions.

In the dispersion-corrected atom-centered potential (DCACP) approach
of von Lilienfeld et al.\cite{vonLilienfeld2004}, van der Waals
interactions are handled by means of an effective electron-core
interaction. Typical plane-wave density functional theory approaches
utilize pseudopotentials, which treat nuclei and core electrons
together as an effective, angular momentum-dependent potential. The
potential is designed such that the all-electron wave-function is
reproduced faithfully. In the DCACP approach, the non-local piece of
this effective core potential is optimized to reproduce high-level
calculations of molecular properties, specifically, the dispersion
energies and forces within molecules. Since the DCACP method uses the
same type of effective core potential that is traditionally used in
plane-wave calculations, its use does not impose additional
computational complexity. The effective potential is designed as a van
der Waals correction to standard gradient-corrected
exchange-correlation functionals, so potentials must be optimized for
each type of atom and for every exchange-correlation functional that
the method is to be paired with. Optimized effective potentials have
been generated for all the standard biological atoms
(carbon\cite{vonLilienfeld2004,Lin2007a}, nitrogen\cite{Lin2007a},
oxygen\cite{Lin2007a}, hydrogen\cite{Lin2007a},
sulfur\cite{Aeberhard2009}, and phosphorus\cite{Cascella2009}), each
with several gradient-corrected functionals. The method shows good
transferability and has been used in molecular\cite{Lin2007}
as well as solid-state applications\cite{Aeberhard2009,Cascella2009}.

Although van der Waals interactions are generally thought of as a
\textit{correlation} effect, the approach of Becke and Johnson takes a
wholly different viewpoint, treating them instead as arising from
interactions between an electron-exchange hole pair in one system and
an induced dipole in another\cite{Becke2005}. This viewpoint is
motivated by the fact that the exchange hole is, in general, not
spherically symmetric, so the electron-exchange hole system has a
non-zero dipole moment. This dipole moment does not affect the energy
of the system containing the electron-exchange hole pair since only
the spherical average of the exchange hole enters the energy
expression for a system. This electron-hole dipole can correlate with
a separate system, however, yielding a dispersion-like
interaction. When averaged over the entirety of a system, the approach
yields molecular $C_6$ coefficients in good agreement with those from
high-level methods\cite{Becke2005}. These can be decomposed into
atomic $C_6$ coefficients and used in a scheme similar to that in the
DFT-D approach\cite{Becke2005a}. A required component of the approach
is the dipole moment of the exchange hole, which Becke and Johnson
conveniently cast as a meta-GGA functional by utilizing the
approximate Becke-Roussel form for the exchange
hole\cite{Becke2005b}. Further development led to the ability to
calculate $C_8$ and $C_{10}$ coefficients\cite{Becke2007}. This
approach is simple, elegant, and performs well over a variety of
systems.

\section{Applications to Biochemistry and Molecular Biology}

As can be seen from Fig.~\ref{scaling}, of all the quantum mechanical
methods discussed above, only DFT is currently capable of treating
systems consisting of several hundred to several thousand atoms---i.e.\
the lower end of the range of biologically relevant molecules. As such,
in this application section we will almost exclusively focus on studies
that have used DFT to investigate such systems.

The methods outlined above represent current state-of-the-art DFT as it
applies to vdW-rich systems. In what follows, these methods' ability to
do useful biochemistry will be highlighted through a brief survey of
recent studies conducted both to test them and to learn from them.
\textit{This survey is intended to act as a showcase of the capabilities
of modern DFT, rather than a comparison of particular methods of its
implementation.}

\subsection{Small molecules}

Small molecules make a natural proving ground for new methods in DFT
because calculations can be compared with quantum chemistry methods.
There exists an extensive body of work, much of it carried out by
\v{S}poner and Hobza\cite{Hobza2008,Jurecka2006} and, independently, by
Stefan Grimme\cite{Antony2006}, benchmarking the DFT methods discussed
above against accurate wave function approaches with special focus being
placed on biologically-relevant molecular systems. This work has yielded
encouraging results and forms the foundation upon which studies of the
physics in these systems rests. But studying small molecules is useful
in its own right, since these play a pivotal role in biochemistry. Most
notable among the biologically-relevant small molecules are water and
the building blocks of macromolecules themselves, namely, DNA bases and
amino acids.

\subsection{Water}

Water has received special attention in the literature, both because of
its great importance to (bio)chemistry and because an accurate
first-principles understanding of it has proven surprisingly elusive.
Most molecular interactions within living systems occur in an aqueous
environment, so an understanding of water is a necessary precursor to
developing an understanding of \emph{in vivo} biochemistry.

These days, the bulk behavior of water (e.g.\ phase diagram, radial
distribution functions) is well modeled by parameterized force
fields.\cite{Jorgensen2005,Mahoney2000,Sanz2004} Although these force
fields get many of the properties of water correct compared with
experiment, the fact that they were parameterized to do just that limits
their usefulness as a tool for understanding the atomistic interactions
in water. At the fundamental level there are quantum effects, most
notably the quantum-mechanical nature of the hydrogen
nuclei,\cite{Vega2010,Li2011b} that cannot be easily reproduced with
classical models. This clouds the connection between microscopic effects
and bulk behavior. A full understanding of the behavior of water can
only come from a quantum mechanical description that applies at the
microscopic level, but can be extended up to the macroscopic limit. 

The behavior of small water clusters (H$_{2}$O)$_{n}$ with $n$ less than
about 6 has been extensively studied at the quantum level and is largely
understood.\cite{Liu1996,Gregory1997,Xantheas2002,Santra2007,Dunn2006,Xantheas1993,Xantheas1994}
Minimum energy geometries can be calculated with high level wave
function methods and these have been compared with various DFT
treatments.  At this level, standard DFT does a reasonable job at
describing the geometric and energetic properties of water, but some
improvement can be made by including dispersion
interactions.\cite{Jonchiere2011,Lin2009,Santra2011} Although the
hydrogen bonds that govern water's structure are not typically thought
of as a van der Waals effect, recent studies have shown that geometries,
energies, dipole moments, and vibrational frequencies of small water
clusters are all improved by inclusion of van der Waals
interactions,\cite{Kolb2011,Silvestrelli2009,Wang2011} as can be seen in
Figs.~\ref{Flo:water_properties_1} and \ref{Flo:water_properties_2}.

\begin{figurehere}
\vspace{4ex}
\centerline{\includegraphics[width=\columnwidth]{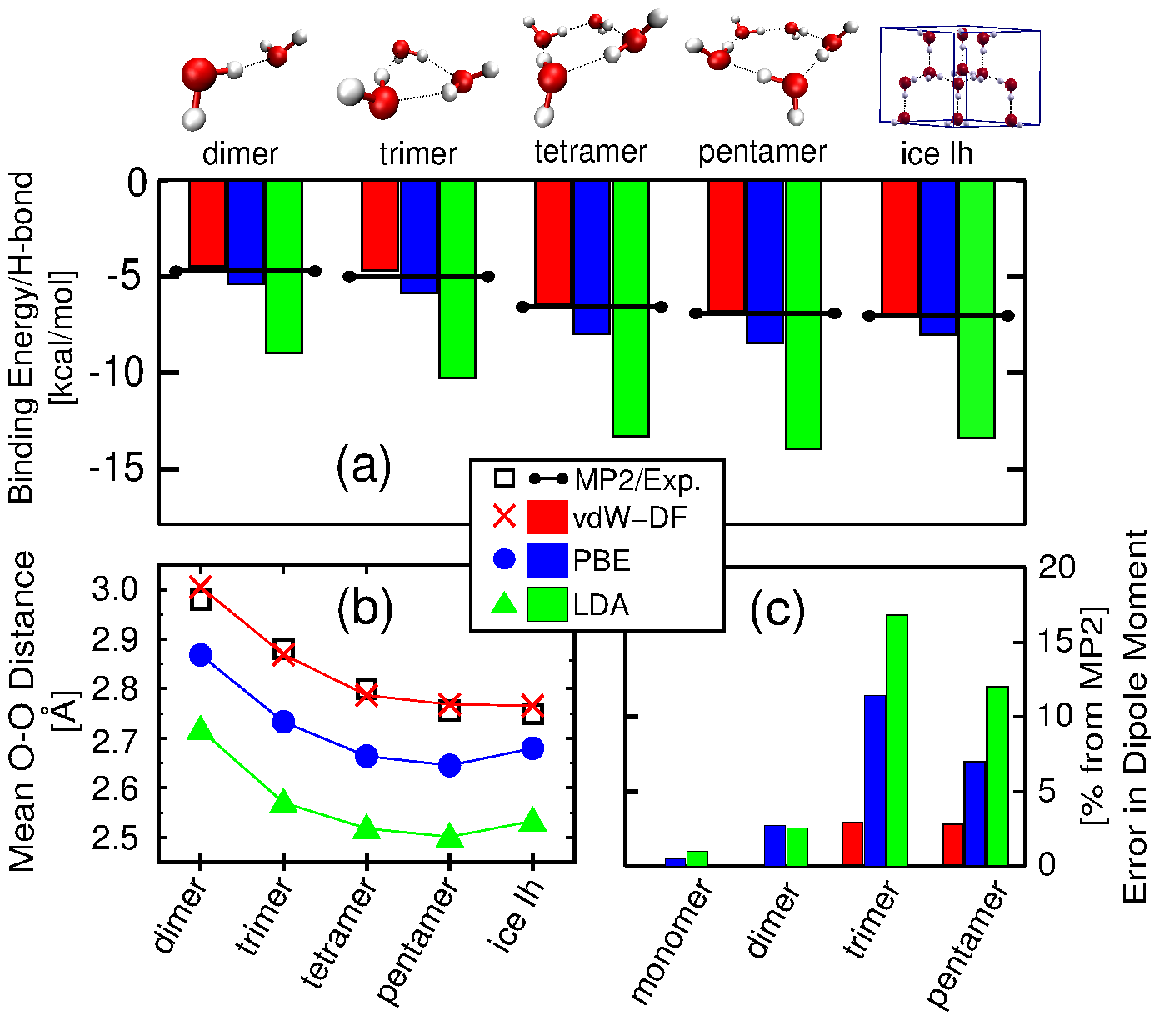}}
\caption{\label{Flo:water_properties_1}Systematic improvement in the
description of water properties with the inclusion of van der Waals
interactions. Calculations on small water clusters including van der
Waals interactions (vdW-DF) compared with standard local (LDA) and
gradient-corrected (PBE) functionals.  Shown are the binding energies,
equilibrium geometries, and dipole moments for each cluster.  (Reprinted
with permission from Ref.~[\protect\refcite{Kolb2011}]; \copyright\ 2011
American Physical Society).}
\end{figurehere}

The improved description of water when van der Waals effects are
included is not limited to small water clusters, but continues into
the bulk\cite{Lin2009,Kolb2011,Jonchiere2011,Paesani2009}.  Through a
series of \emph{ab initio }molecular dynamics simulations Lin et
al.\cite{Lin2009} showed that the radial distribution functions
produced by standard gradient-corrected functionals tend to produce
water that is over-structured compared with experiment. This was also
evident in the average number of hydrogen bonds and the self-diffusion
coefficient, both of which show an over-structuring of the water
molecules.  These results mirror obtained by numerous other groups
working with a variety of different codes, exchange-correlation
functionals, and basis
sets\cite{Serra2005,Kuhne2009,Grossman2004,Schwegler2004,Kumar2007} This
over-structuring is mitigated to a large degree by a proper treatment
of van der Waals interactions. The self diffusion coefficient
increases three-fold and the over-structuring evident in the radial
distribution functions softens when van der Waals interactions are
included. This is also true for bulk ice in its standard hexagonal
form (I$_{h}$) where inclusion of van der Waals interactions again
improves the description of structural and electronic properties.

\begin{figurehere}
\vspace{4ex}
\centerline{\includegraphics[width=0.85\columnwidth]{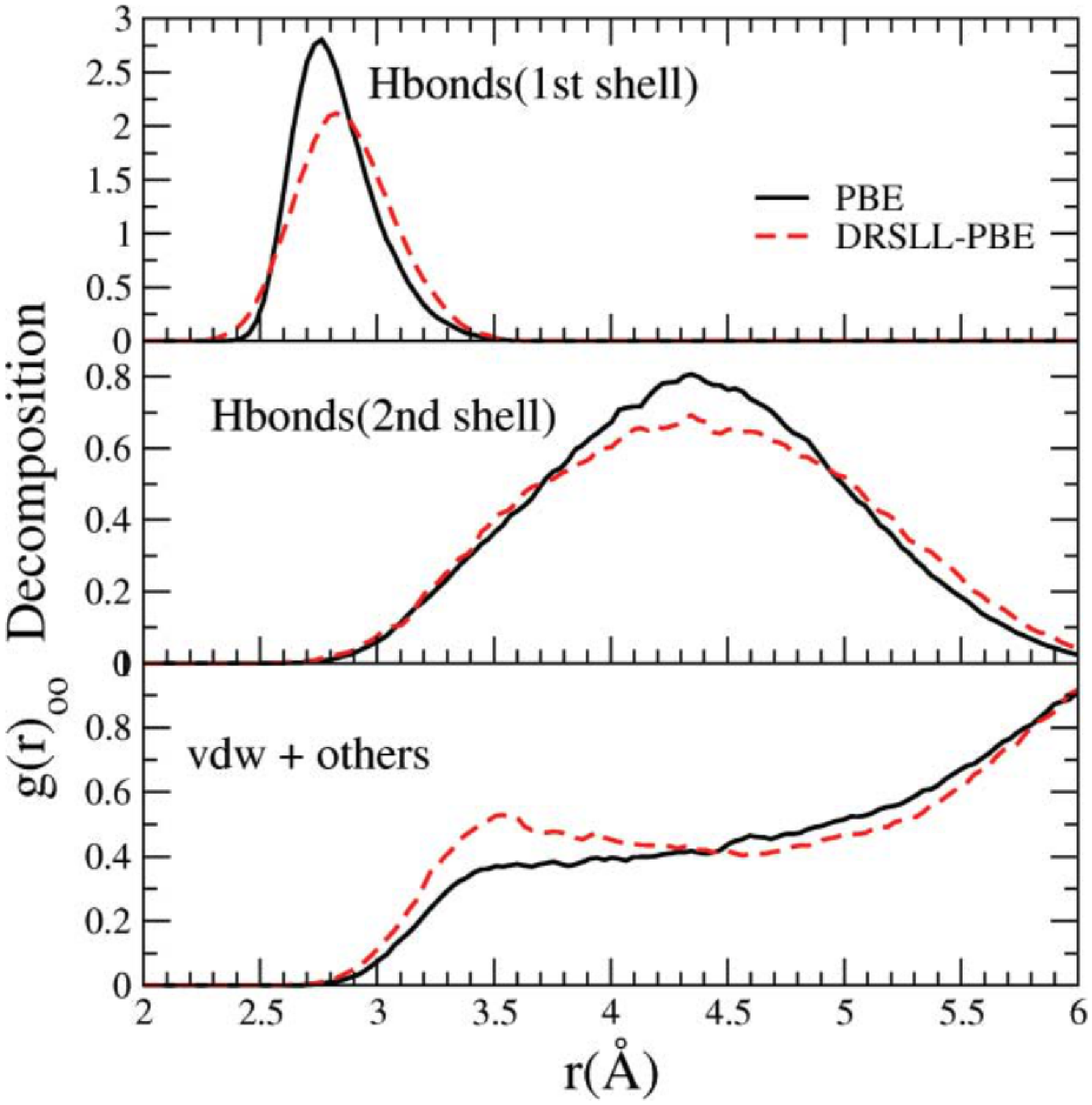}}
\caption{\label{Flo:water_properties_2}Decomposition of the
oxygen-oxygen radial distribution function as calculated by a standard
gradient-corrected functional (PBE) and vdW-DF (here called DRSLL-PBE).
The interactions are broken into (from top to bottom) first coordination
shell hydrogen bonds, second coordination shell hydrogen bonds, and
higher-order interactions. (Reprinted with permission from
Ref.~[\protect\refcite{Wang2011}]; \copyright\ 2011 American Institute
of Physics).}
\end{figurehere}

It is worth pointing out that the results of DFT calculations can vary
quite widely depending on the choice of basis set and
exchange-correlation functional used, and great care must be taken
with their selection. For example, when coupled to the non-local piece
of the vdW-DF, the overly repulsive revised PBE exchange functional
actually produces water that is \textit{under-structured} compared
with experiment, in contrast to most other exchange functionals. This
is related to the aforementioned tendency of the original vdW-DF to
predict intermolecular interaction distances that are large compared
with experiment and high-level wave-function methods. Additionally, it
has been pointed out that the properties of liquid water calculated
within DFT can depend quite strongly on the choice of basis
set.\cite{Kumar2007} Calculations similar to those shown in
Fig.~\ref{Flo:water_properties_2} were carried out by Zhang et al. and
showed considerably less improvement in the oxygen-oxygen radial
distribution function compared to experiment\cite{Zhang2011}. The
basis sets used in the two sets of calculations were fundamentally
different, making a direct comparison of their appropriateness
difficult. Despite these issues, it is generally agreed that, when
properly chosen basis sets and exchange-correlation functionals are
used, the inclusion of van der Waals interactions fundamentally
improves the DFT description of water, both at the microscopic level
and in bulk.

It is interesting to note that, although inclusion of van der Waals
interactions greatly improves the description of water, this alone
does not complete the picture of important effects within water. The
standard Born-Oppenheimer approximation used in quantum-mechanical
studies treats all nuclei as classical point particles. Recent work by
a number of groups, however, has shown that nuclear quantum effects
may play a significant role in determining the properties of
water\cite{Fanourgakis2006,Morrone2008,Chen2003,Li2011b}. In fact, it
has been shown that such nuclear quantum effects may be more
far-reaching, playing a substantial role in hydrogen bonds in general,
not just between water molecules\cite{Li2011b}. This would, of course,
have enormous consequences for a proper description of interactions
within biological molecules such as proteins and DNA, where hydrogen
bonds often dominate the binding. For example, it has been
proposed\cite{Lowdin1963} that the keto form of DNA nucleobases (the
standard form required for Watson-Crick hydrogen bonding) can
spontaneously tautomerize via hydrogen tunneling to the enol form, a
process which could be responsible for some types of DNA damage. A
recent study by P\'{e}rez et al. found that, although such tunneling
does occur, the metastable enol form has a lifetime too short to play
a significant role in DNA mismatch damage\cite{Perez2010}. In fact,
the effects of quantum nuclei appear to dynamically stabilize the keto
form.  For a recent review of these considerations see
Ref. \refcite{Pu2006}.

\subsection{DNA nucleobases}

The four nucleobases, arranged in different sequences along strands of a
sugar-phosphate polymer, have enabled the information of life to be
stored and propagated since life began. Each of these relatively simple
molecules contains an aromatic ring capable of engaging in multiple
hydrogen bonds. When these bases come together in a Watson-Crick,
\textit{edge-on} manner, they can form hydrogen bonds strong enough to
hold two DNA strands together. When brought together in a parallel
\textit{face-on} fashion, they form $\pi$--$\pi$ stacking interactions
strong enough to give it an average persistence length of roughly 50 nm
with some sequences having even larger persistence
lengths\cite{Bednar1995}.

\begin{figure*}
\centerline{\includegraphics[width=1.6\columnwidth]{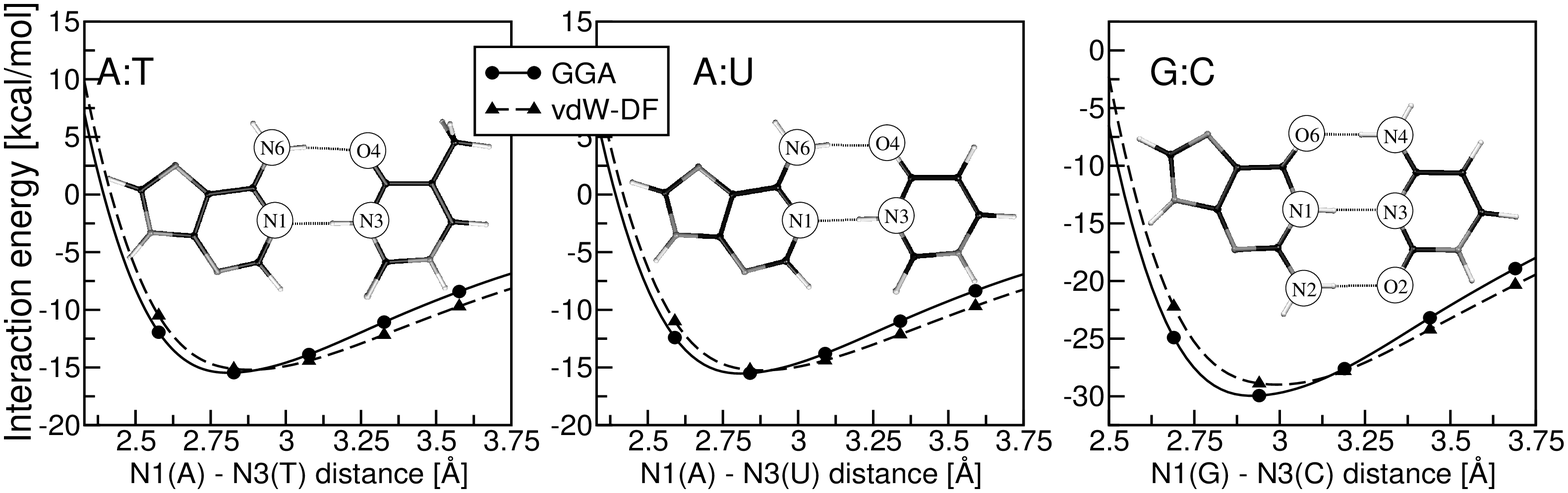}}
\caption{\label{fig_h-bond} Interaction within A:T, A:U, and G:C base
pairs as a function of N1$\cdot$$\cdot$$\cdot$N3 separation for the
canonical Watson-Crick, edge-on orientation.  Both GGA (solid circles)
and vdW-DF (solid triangles) energies are shown. The labeled atoms are
the non-hydrogen atoms engaging in hydrogen bonding.  (Reprinted with
permission from Ref.~[\protect\refcite{Cooper2008a}]; \copyright\ 2008
American Institute of Physics).}
\end{figure*}

\begin{figure*}
\begin{center}
\includegraphics[width=1.65\columnwidth]{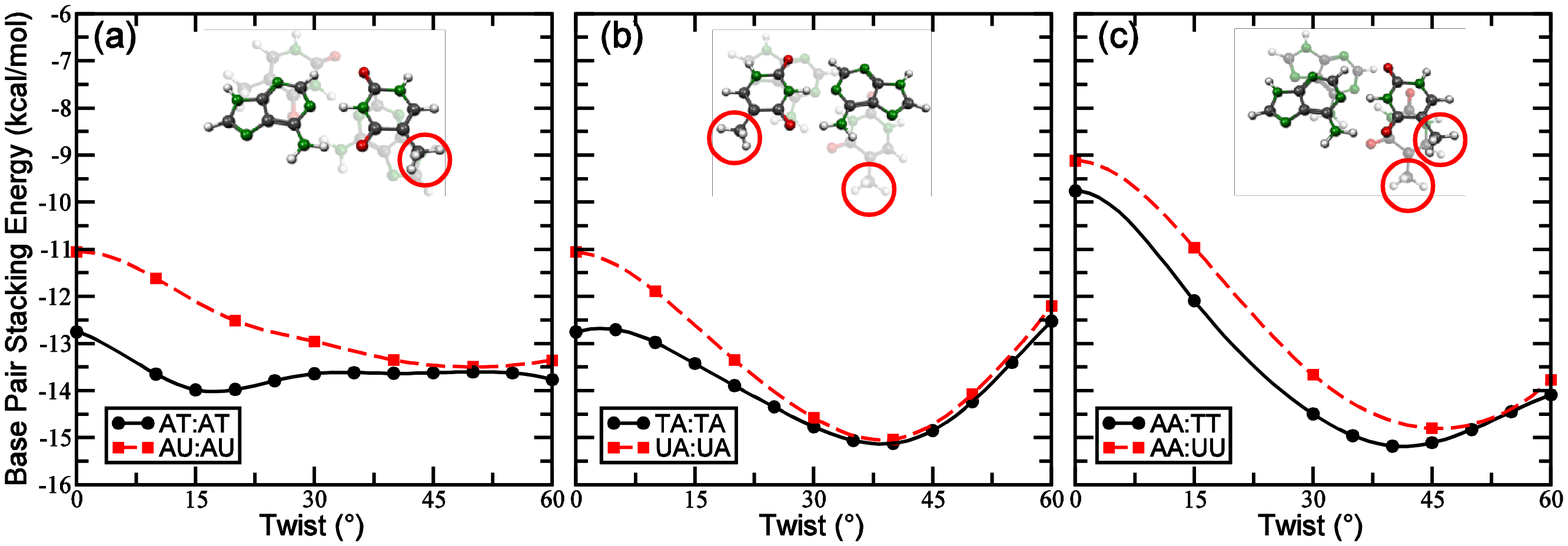}\\
\includegraphics[width=1.65\columnwidth]{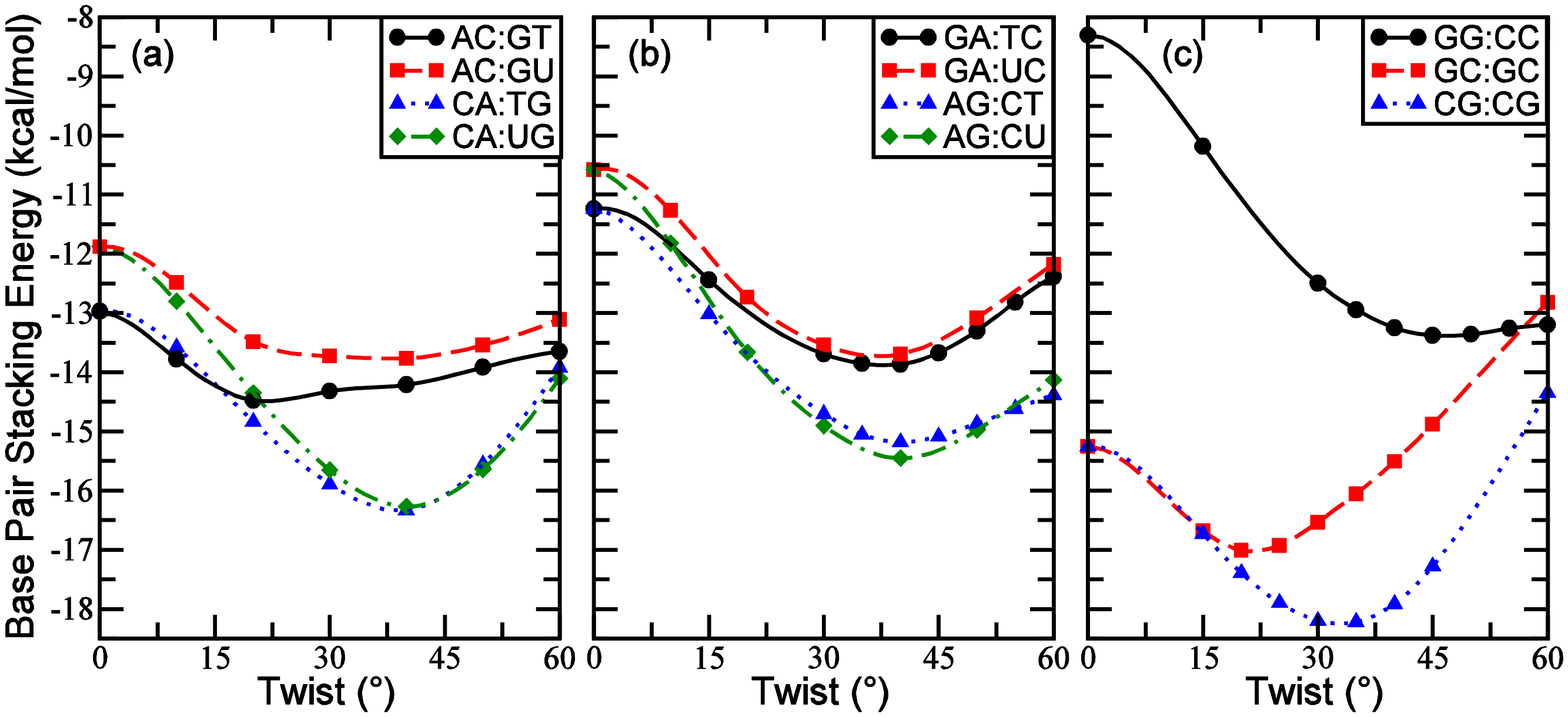}
\end{center}
\caption{\label{fig_stacking} Interaction energy as a function of twist
for all possible DNA base pair steps as well as those for their
uracil-containing counterparts.  The insets give a top view of the
molecular interaction for the AT:AT (top left), TA:TA (top center), and
AA:TT (top right) steps.  All calculations were carried out using the
vdW-DF.  (Reprinted with permission from
Ref.~[\protect\refcite{Cooper2008}]; \copyright\ 2008 American Chemical
Society).}
\end{figure*}

Cooper, Thonhauser, and Langreth calculated the base interaction energy
as a function of distance for a Watson-Crick, \textit{edge-on} approach
of two base pairs (see Fig.~\ref{fig_h-bond}).\cite{Cooper2008a} This
was done for the A:T, A:U, and G:C combinations. The G:C base pair
exhibits a maximum interaction energy of about twice that of the other
pairs, not surprising since it has an extra hydrogen bond, and all three
show similar equilibrium binding distances.

The base stacking energy as a function of geometry has been studied by
several groups.\cite{Fiethen2008} The binding energies as a function of
twist angle for all possible stacked base pairs are shown in
Fig.~\ref{fig_stacking}. It is noted that the methyl substitution that
differentiates thymine from uracil stabilizes the systems with respect
to twist.

In 2006, Jurecka et al. published a set of accurate, quantum-chemical
binding energies of 22 molecular dimers, selected for the importance of
van der Waals interactions within them.\cite{Jurecka2006} The set
(dubbed the S22 dataset) was broken into three distinct groups: (i)
dimers for which hydrogen bonding is the key component of binding, (ii)
dimers for which pure dispersion is the key component of binding, and
(iii) dimers which exhibit a mixture of both of these effects.
Comparison with this dataset became the \textit{de facto} metric for
assessing the ability of fledgling methods within DFT to correctly
account for van der Waals interactions.  Within this set (which was
later revised, expanded, and placed in a convenient online
database)\cite{rezac2008} were a homo\-dimer of uracil and an A:T
heterodimer.

\begin{figurehere}
\vspace{4ex}
\centerline{\includegraphics[width=0.9\columnwidth]{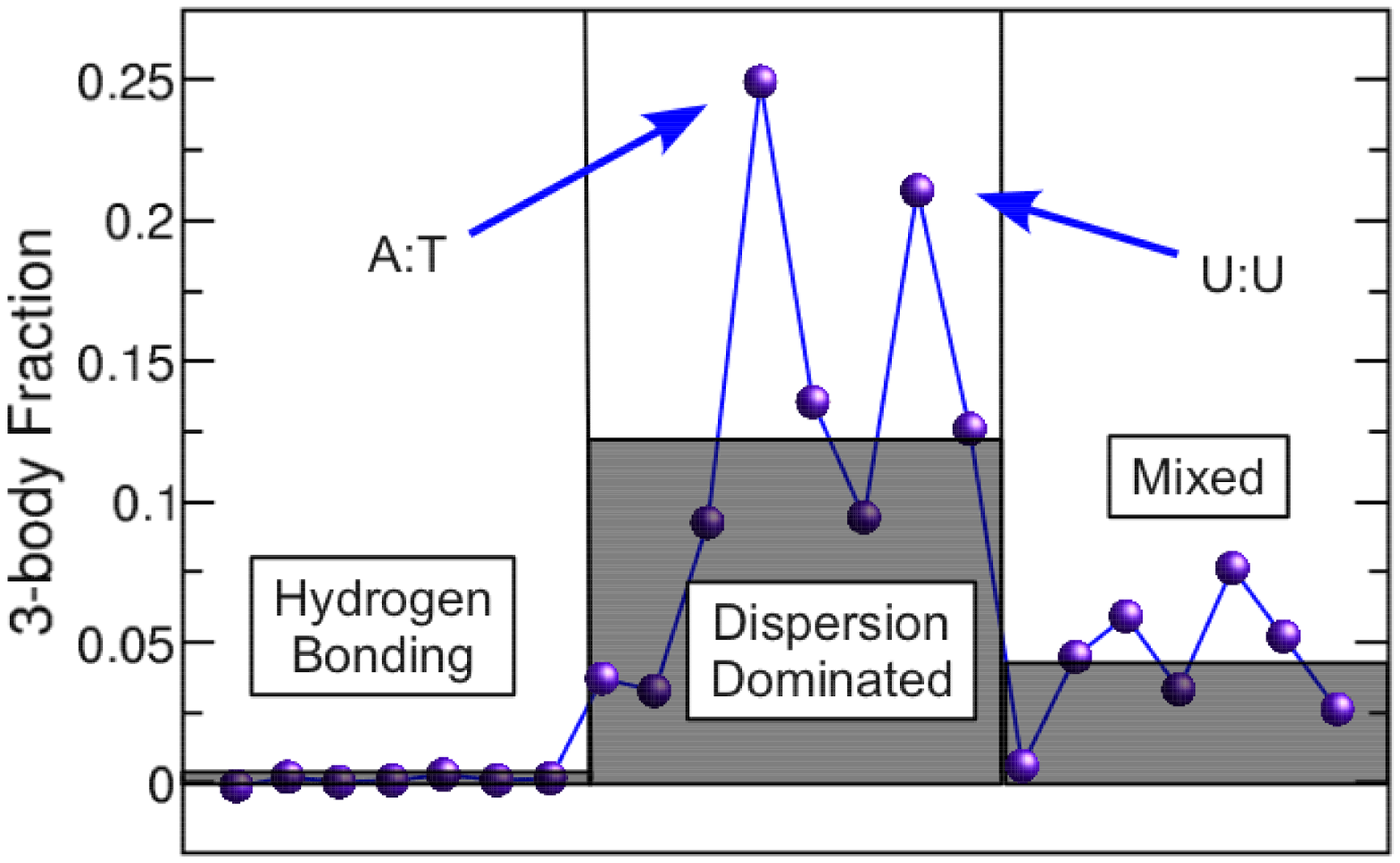}}
\caption{\label{3-body_s22}Ratio of 3-body dispersion term to total
CCSD(T) binding energy for the 22 systems in the S22 data set (data
taken from Ref.~\protect\refcite{VonLilienfeld2010}). Note that the
hydrogen-bond dominated systems show little dependence on 3-body
dispersion while the dispersion systems show variable but significant
3-body contributions. Specifically marked are the uracil-uracil U:U
and adenine-thymine A:T dimers, which exhibit the highest 3-body
dispersion fraction of all the systems. The gray boxes denote the
average 3-body dispersion fraction within each group.}
\end{figurehere}

In 2010, a landmark paper by von Lilienfeld and Tkatchenko showed that
the uracil-uracil U:U and adenine-thymine A:T stacked bases exhibit
large 3-body dispersion terms\cite{VonLilienfeld2010}. Going a step
further, the authors addressed the magnitude of two and three-body
dispersion interactions across the entire S22 dataset. Some of their
results are shown in Fig.~\ref{3-body_s22}. Using the DFT+vdW approach
enhanced with the triple-dipole term (as discussed in Section
\ref{sec_methods}), they found that the three distinct groups of the S22
set show markedly different dependencies on three-body dispersion
interactions. The systems showing large 3-body dispersion terms (which
include the stacked U:U and A:T dimers) were the systems deemed
dispersion-dominant and those with essentially no 3-body dispersion
interactions were systems dominated by hydrogen bonding. The authors
argue that 3-body effects may be more important than previously thought.
Interestingly, for stacked nucleobases the 3-body dispersion term seems
to be relatively constant, especially compared with the pairwise
dispersion term. Figure~\ref{3-body_all} shows the 2 and 3 body
dispersion terms calculated by von Lilienfeld and Tkatchenko for 42
stacked nucleobases and base pairs. The 3-body contribution to the
energy is relatively constant across the entire dataset while the 2-body
term varies considerably, especially for the weaker-binding systems.

\begin{figurehere}
\vspace{4ex}
\centerline{\includegraphics[width=0.9\columnwidth]
{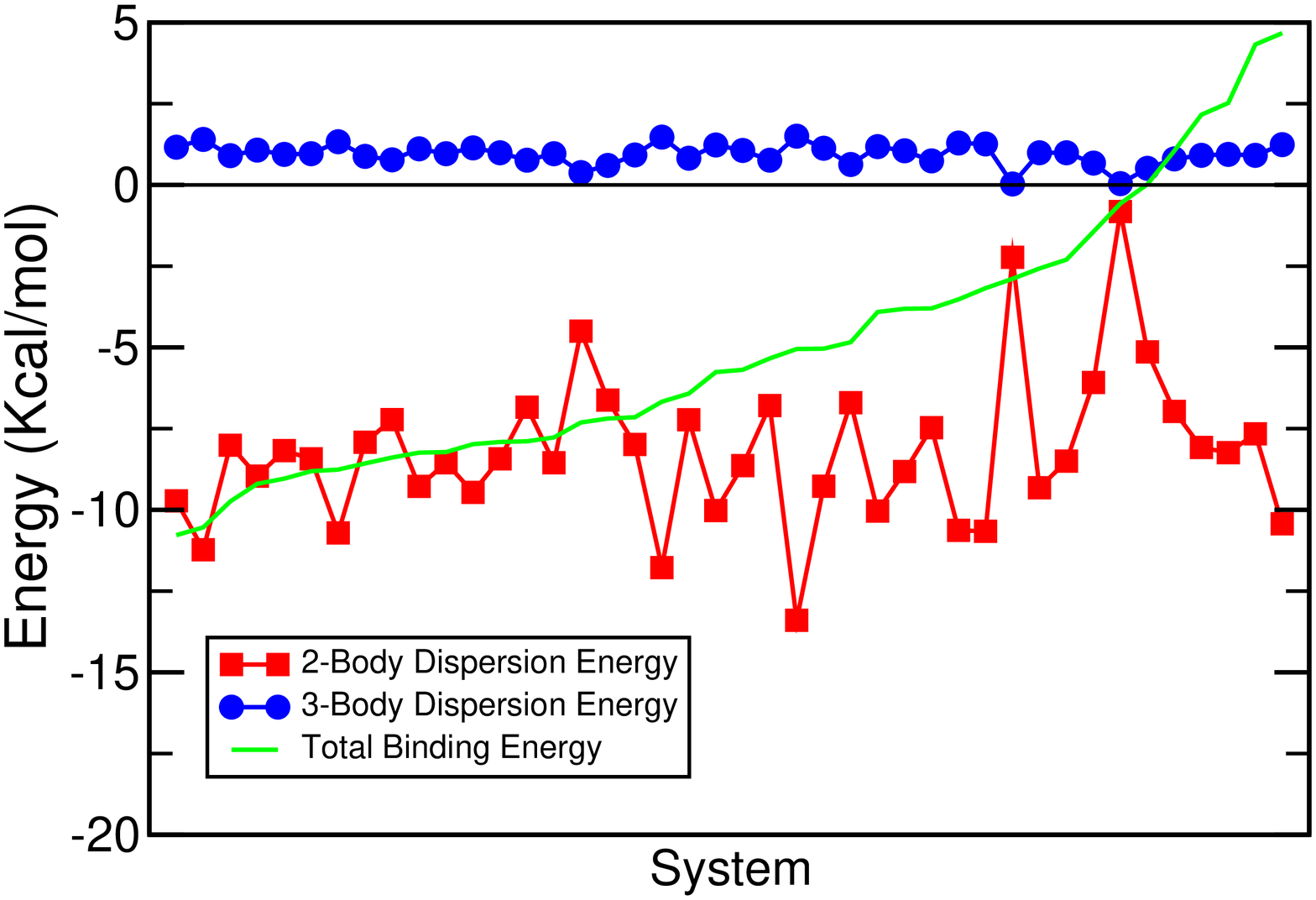}}
\caption{\label{3-body_all} Contributions from 2-body (red boxes) and
3-body (blue circles) dispersion interactions over all stacked base
pairs investigated by \vl and \tk (data taken from
Ref.~\protect\refcite{VonLilienfeld2010}). The green line shows the
total binding energy as calculated by CCSD(T).}
\end{figurehere}

\subsection{DNA intercalation}

The $\pi$--$\pi$ stacking interactions of DNA bases discussed in the
previous section are important for another reason. Many cancer-causing
agents act by intercalating between base-pairs within a strand of DNA,
preventing it from carrying out its normal functions. Ironically, some
anti-cancer drugs can also act in this way. In the latter case the DNA
is intentionally disturbed either to prevent its replication or to
trigger cell death.

One well known intercalating anticancer drug is the poly-aromatic
ellipticine molecule. This molecule can intercalate between base pairs
of DNA where it is believed to interfere with the process of
replication, effectively killing the cell.\cite{Auclair1987} Li et al.
calculated the binding energy between the neutral ellipticine molecule
and a single C:G base pair to be --18.4 kcal/mol.\cite{Li2009} Not
surprisingly, the strength of the binding was shown to have a
substantial dependence on the relative angle between the ellipticine and
DNA bases, showing a relatively strong (several kcal/mol) preference for
near parallel and anti-parallel conformations. Chun Lin et al.
investigated the intercalation of ellipticine between a cytosine-guanine
base step (i.e. a pair of C:G base pairs).\cite{Lin2007} As shown in
Fig. \ref{ellipticine_intercalation}, they found that ellipticine is
significantly attracted to the DNA complex even when it is several
angstroms away and ultimately intercalates with a binding energy of
about 37 kcal/mol, in perfect accord with the earlier results found by
Li et al. Further, they found that the interaction was repulsive when
van der Waals interactions were excluded from the calculations. von
Lilienfeld and Tkatchenko found that the pairwise dispersion energy for
this system to be a substantial \mbox{--57}~kcal/mol and the 3-body
correction term was 8.9 kcal/mol.\cite{VonLilienfeld2010} This is
certainly a significant binding event and shows the strength with which
aromatic molecules can interact with the loose $\pi$ electrons within
DNA. Such interactions are common for $\pi$-stacked molecules such as
benzene and the large number of relatively close neighbors within the
$\pi$-conjugated DNA bases is believed to be responsible for the large
interaction energy.

\begin{figurehere}
\vspace{4ex}
\centerline{\includegraphics[width=0.9\columnwidth]{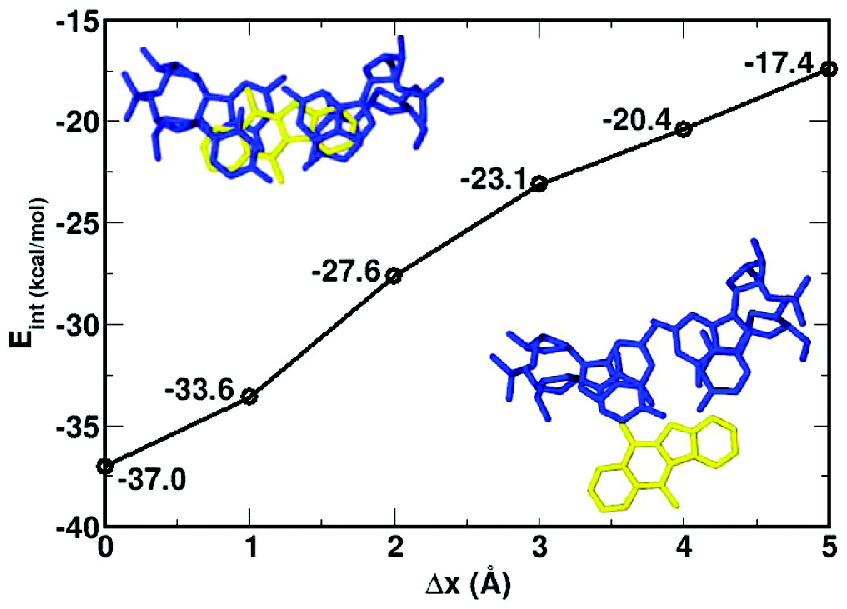}}
\caption{Interaction energy of an ellipticine molecule as it
intercalates into a CG:CG base step.  Here $\Delta x$ represents the
displacement of the ellipticine from the final intercalated position.
The insets correspond to the fully intercalated structure at $\Delta
x=0$ \AA\ (top left) and $\Delta x=5$ \AA (bottom right).  (Reprinted
with permission from Ref.~[\protect\refcite{Lin2007}]; \copyright\ 2007
American Chemical Society).}
\label{ellipticine_intercalation}
\end{figurehere}

The intercalation of both positively charged and neutral proflavine has
also been studied.\cite{Li2009} Neutral proflavine was found to bind to
a C:G base pair with an energy of about \mbox{--20.3}~kcal/mol and
charged proflavine with an energy near 12.1 kcal/mol. The difference
between the charged and uncharged binding was attributed to
electrostatic effects rather than those of correlation. This conclusion
was reached largely because the results of standard PBE calculations,
although they get the interaction energy of each system wrong, exhibit a
similar \emph{difference }between the two binding energies. It is
interesting to note that the binding energy is larger for the positively
charged proflavine even though the negatively charged backbone was
omitted from these calculations. Again, a substantial preference for
near parallel and anti-parallel relative angles was found. 

The energetics of the interaction between proflavine and a T:A base pair
was also studied, with results qualitatively similar to those of the
proflavine--C:G system.\cite{Li2009} Proflavine was found to bind to T:A
with a binding energy of --18~kcal/mol, again showing preference for
near parallel and anti-parallel configurations. Steric clashes with the
methyl group on the thymine base produced some interesting features in
the rotation curve but did not change the overall preferred structures.

Perhaps the most interesting finding of Li et al. was that both
intercalators studied were found to have stronger interactions with a
C:G base pair than a C:G base pair has with \emph{another} C:G base
pair, and that the angular dependence of these interactions
qualitatively differ.\cite{Li2009} A C:G base pair dimer has a
double-well minimum centered around 0$^\circ$, with the minimum-energy
configurations at a twist of about 35$^\circ$ and --35$^\circ$. The
intercalators, by contrast, exhibit only one of these minima.  This may
partially explain the disruption of secondary structure observed upon
intercalation of these molecules, which may play an important role in
their anti-cancer function.\cite{Berman1981}

\subsection{Proteins}

Owing to their large size and complexity, simulation of proteins often
proves to be a formidable challenge even for simple parameterized
models. The application of quantum mechanics to a full protein is,
unfortunately, still beyond the reach of modern DFT. Recently, however,
significant steps toward a quantum understanding of proteins have been
made.

Helical chains of alanine molecules are often studied because they are
relatively simple yet they exhibit the canonical helix structure present
in so many proteins. In addition, when capped with a charged species
they can be formed experimentally so computed properties may be compared
with experiment.\cite{Hudgins1998} In one study, Tkatchenko et al.\
looked at three helical forms ($\alpha$, $\pi$, and 3$_{10}$) of
poly-alanine chains.\cite{Tkatchenko2011} By comparing with PBE, a
standard gradient-corrected functional, they found significant van der
Waals stabilization of all three helix types relative to the fully
extended structure. In fact, PBE predicts nearly equal stabilization
energies for all three whereas the van der Waals calculations showed a
splitting of about 2 kcal/mol between the $\alpha$-helix and 3$_{10}$
structures. The authors note that the van der Waals effects are of much
shorter range than the standard hydrogen bond stabilizations in the
helical forms, since the helices are long-ranged structures exhibiting
periodic hydrogen bonds. Despite this, the study found that van der
Waals interactions were critical to explain the observed stability of
poly-alanine helices up to about 700 K. Through \emph{ab initio}
molecular dynamics calculations Tkatchenko et al. found that, when van
der Waals effects were excluded, the helical structure gave way to the
fully extended form at a temperature well below that observed
experimentally, even though hydrogen bonds were still correctly
accounted for. Agreement with experiment was recovered when van der
Waals interactions were included in the calculations, which showed a
breaking up of the helical structure between 700 and 800 K.

Drug discovery is a multi-billion dollar business and much effort is
being put into so called \emph{rational drug design} where potential
drug molecules are scored based on their predicted binding affinity to a
particular protein target.\cite{Gohlke2001,Verlinde1994,Abagyan2001}
This transfers the \emph{trial-and-error }phase away from the lab, where
experiments to test drug binding affinities can be relatively expensive
and time-consuming, to the computer, where thousands of potential drugs
can be tested for binding at relatively low
cost.\cite{Abagyan2001,Gane2000} Working toward this end, Antony et al.
studied the interactions of a number of protein active sites with their
respective biological ligands.\cite{Antony2011} They found that
exclusion of van der Waals interactions can substantially change the
ordering of ligand binding affinities. Further, they found that neglect
of these interactions can actually lead to the computed binding energies
for a ligand with its target receptor being of the wrong sign.

Another study, carried out by Rutledge and Wetmore focused on ligands
that interact with their host protein via $\pi$--$\pi$ stacking
interactions and T-shaped $\pi$ interactions.\cite{Rutledge2010} As
before, they found that inclusion of van der Waals effects is imperative
to obtain accurate energetics in such systems.

\section{Future Directions}

With the utility of these methods established, attention can be turned
toward the future and what can be accomplished with them. Computation of
a full macromolecule in atomistic detail is still beyond the reach of
DFT, even for the most advanced computers, but the method can still be
used as a tool to aid in our understanding of such systems. 

One useful approach that has been adopted by some groups is to use DFT
to parameterize new force fields. Typically, these are parameterized
either to reproduce experimental results or the results of high-level
quantum chemistry calculations. As discussed in Section
\ref{sec_methods}, quantum chemistry methods are limited to fairly small
systems. Parameterizing force fields using the much larger systems that
DFT is capable of simulating might help average out size effects and
better represent the environment that exists within macromolecules.
Additionally, solid-state parameter sets could be developed to deal with
molecular crystals.

Another useful application of DFT is in the refinement of experimental
structures. Typical x-ray and NMR techniques provide data that is
consistent with more than one structure. Also problematic is the
placement of the x-ray invisible hydrogen atoms. Given this,
experimentalists often use semi-empirical calculations to refine the
observed structure.  Use of high-level DFT calculations including van
der Waals interactions could yield a better result, since large systems
can be calculated very accurately.

Drug discovery is another area where useful progress is being made by
incorporating DFT calculations and it is expected that DFT will play an
important role in this area soon.\cite{Patel_99} Although an entire
protein may not be able to be treated quantum mechanically, hybrid
methods that apply varying levels of theory to regions within a protein
are being used with much success. One can treat the drug molecule and
its binding site with full quantum-mechanical rigor while treating
distant regions using well-tested classical or semi-empirical
approaches. This allows the most important physics to be treated
accurately and coupled to a sufficient treatment of the less important
parts of the problem.  This is not only a useful approach for drug
design but also applies to understanding the normal operation of
ligand-binding proteins.  Such methods are general referred to as QM/MM,
i.e.\ quantum mechanics/molecular mechanics.

Finally, although the applicability of DFT to calculations on full
macromolecules is currently limited, linear scaling DFT methods are
becoming popular and provide a tantalizing way forward. These
approaches, which make use of special algorithms and highly-localized
basis functions, can easily treat thousands of atoms---see
Fig.~\ref{scaling}. Such capabilities make computation of full
macromolecular systems feasible. For example, the fledgling
linear-scaling code ONETEP has been used to calculate properties of a 20
base-pair strand of DNA containing almost 1300 atoms. If augmented with
the ability to adequately treat dispersion interactions, such linear
scaling DFT approaches may provide a practical means to apply full
quantum mechanics to biological problems of real interest in the
near future.


\end{multicols}
\end{document}